\documentclass[prd,twocolumn,amssymb,floatfix,preprintnumbers,nofootinbib]{revtex4}
\usepackage{graphicx,amsmath,mathtools}
\usepackage{epsfig,subfigure}
\usepackage{dcolumn}
\usepackage{bm}
\usepackage{float}
\usepackage{multirow}

\def\beq{\begin{equation}}
\def\eeq{\end{equation}}
\def\bea{\begin{eqnarray}}
\def\eea{\end{eqnarray}}

\def\3Eqs#1#2#3{Eqs.\ (\ref{#1}), (\ref{#2}) and (\ref{#3})}

\def\within#1to#2/{#1\mbox{ to }#2}
\def\lbullet#1,#2,#3,#4/{\Text(#1,#2)[c]{$\bullet$}\Line(#1,#2)(#3,#4)}
\def\textcite#1{Ref.~\cite{#1}}

\def\newu{$U(1)_{L_\mu-L_\tau}$}

\def\hd0ino{\tilde{h^0_{d}}}

\def\hu0ino{\tilde{h^0_{u}}}

\def\e2spl{\ensuremath{^\clubsuit}}

\def\mE{E \hspace{-.7em}/\;\:}
\def\gsim{\lower0.5ex\hbox{$\:\buildrel >\over\sim\:$}}
\def\lsim{\lower0.5ex\hbox{$\:\buildrel <\over\sim\:$}}

\def\nn{\nonumber\\*}

\begin{document}

\title{Signatures of supersymmetry and a $L_\mu - L_\tau$ gauge boson at Belle-II}

\author{Heerak Banerjee} 
\email{tphb@iacs.res.in}
\author{Sourov Roy}
\email{tpsr@iacs.res.in}
\affiliation{School of Physical Sciences, Indian Association for the Cultivation of Science,
2A $\&$ 2B Raja S.C. Mullick Road, Kolkata 700 032, India}

\date{\today}

\begin{abstract}
We propose that the $\gamma + \mE$ signal at the Belle-II detector will be a smoking gun for supersymmetry (SUSY) in the presence of a gauged \newu symmetry. A striking 
consequence of breaking the enhanced symmetry appearing in the limit of degenerate (s)leptons is the nondecoupling of the radiative contribution of heavy charged sleptons 
to the $\gamma - Z^\prime$ kinetic mixing. The signal process, $e^+ e^- \rightarrow \gamma Z^\prime \rightarrow \gamma + \mE$, is an outcome of this ubiquitous feature. 
We take into account the severe constraints on gauged \newu models by several low-energy observables and show that any significant excess in all but the highest 
photon energy bin would be an undeniable signature of such heavy scalar fields in SUSY coupling to $Z^\prime$. The number of signal events depends crucially on the 
logarithm of the ratio of stau to smuon mass in the presence of SUSY. In addition, the number is also inversely proportional to the $e^+-e^-$ collision energy, making a 
low-energy, high-luminosity collider like Belle-II an ideal testing ground for this channel. This process can probe large swathes of the slepton mass ratio vs the 
additional gauge coupling ($g_X$) parameter space. More importantly, it can explore the narrow slice of $M_{Z^{\prime}}-g_X$ parameter space still allowed in gauged \newu 
models for superheavy sparticles.
\end{abstract}

\maketitle

\section{Introduction}
The standard model (SM) of particle physics fails to explain many experimental observations like neutrino mass, presence of dark matter, the baryon asymmetry of the 
universe and the anomalous magnetic moment of muon among others. There is little doubt that we need physics beyond the standard model to address 
these issues. One of the most celebrated extensions of the SM is its minimal supersymmetric extension, popularly called MSSM \cite{Martin:1997ns}. 
However, MSSM with R-parity conservation still cannot explain the nonzero tiny masses of the neutrinos and their nontrivial mixing pattern.

At the same time, given the upward trend of the lower bounds on supersymmetry (SUSY) particle masses, it is also becoming increasingly difficult to accommodate the results from the
muon $(g-2)$ experiment in MSSM. Supersymmetric particle searches from the 35.9 fb$^{-1}$ data, collected by CMS experimental collaborations at the LHC at a 
center-of-mass energy, $\sqrt{s}$ = 13 TeV, have found no significant excess in signal over the expected SM background. Results from the CMS experiment \cite{CMS:2018wbf} 
extend the gluino mass limit to 2.0 TeV, and the limits on top squark masses reach 1.14 TeV. For simplified models and a massless lightest neutralino, limits on the first 
two generations left-handed slepton masses are set at 450 GeV, whereas the limits on the right-handed slepton masses are set at 330 GeV \cite{CMS:2017mkt}. The searches by 
ATLAS experiment \cite{Aaboud:2017vwy} under similar assumptions set an exclusion limit of 2.03 TeV on the gluino mass and 1.55 TeV  on squark masses (first two generations) 
corresponding to an integrated luminosity of 36.1 fb$^{-1}$ at $\sqrt{s}$ = 13 TeV. Masses up to 500 GeV \cite{Aaboud:2018jiw} are excluded for sleptons assuming three 
generations of mass degenerate sleptons. 

There are other models that address the issue of neutrino masses along with one or more of the observations not in tune with the SM. One class of models that has 
generated considerable interest of late is that with an additional gauged \newu symmetry \cite{He:1990pn,He:1991qd}. It has been implemented to explain the muon anomalous 
magnetic moment \cite{Baek:2001kca,Altmannshofer:2016brv,Ma:2001md}, neutrino masses and mixing \cite{Ma:2001md,Heeck:2011wj,Baek:2015mna,Biswas:2016yan}, 
dark matter \cite{Altmannshofer:2016jzy,Biswas:2016yan,Biswas:2016yjr,Patra:2016shz,Biswas:2017ait,Arcadi:2018tly,Kamada:2018zxi,Foldenauer:2018zrz,Cai:2018imb}, 
Higgs boson flavor violating decays \cite{Crivellin:2015mga,Altmannshofer:2016oaq}, and B-decay 
anomalies \cite{Altmannshofer:2014cfa,Crivellin:2015mga,Altmannshofer:2016jzy,Ko:2017yrd,Baek:2017sew}.     
The possibility of detecting the gauge boson of \newu symmetry at the Belle-II experiment has been discussed \cite{Araki:2017wyg,Kaneta:2016uyt}. 
In addition, constraints on the mass and the coupling of the new gauge boson in this model from neutrino trident production \cite{Altmannshofer:2014pba}, 
neutrino beam experiments \cite{Kaneta:2016uyt}, lepton flavor violating $\tau$ decays \cite{Chen:2017cic} and rare Kaon decays \cite{Ibe:2016dir} severely restrict
its parameter space. The situation in supersymmetric versions of \newu is more relaxed and has been studied in the context of muon anomalous magnetic moment, neutrino 
masses and mixing, charged lepton flavor violating decays \cite{Banerjee:2018eaf}, dark matter \cite{Das:2013jca,Darme:2018hqg}, and flavor anomalies \cite{Darme:2018hqg}. 

In this work, we show that the signature of supersymmetry as well as $L_\mu - L_\tau$ gauge boson can be seen at the Belle-II experiment \cite{Abe:2010gxa} by 
studying the process $e^+e^- \rightarrow \gamma + \mE$, which hinges on the presence of kinetic mixing between the photon and the extra gauge boson. Such a kinetic mixing 
is an unavoidable feature in any model with an additional $U(1)$ gauge symmetry. Here we consider a scenario where this mixing is absent at the tree level but arises 
radiatively at the one-loop level. As a result, the kinetic mixing is not constant but depends on the extra gauge coupling, the mass ratio of smuon and stau defined 
as $r = \frac{m_{\tilde \tau}}{m_{\tilde \mu}}$ and the momentum carried by the photon and the $Z^\prime$. When $r$ is 1, the supersymmetric contribution to the kinetic 
mixing vanishes and the results resemble those for the nonsupersymmetric gauged $U(1)_{L_\mu - L_\tau}$ model. However, when the ratio is greater than or less than 1, the 
supersymmetric contribution could be significant, and the results are distinctly different from those of the nonsupersymmetric model. 

As the results do not depend on the absolute mass scales, the signature of SUSY may be observed at the Belle-II experiment through this channel even when the sparticles 
are extremely massive. On the other hand, when all the SUSY particles are very heavy, the contribution to muon $(g-2)$ at one nloop comes only from the loop involving the
$Z^\prime$ gauge boson. The allowed region of parameter space in the $(M_{Z^\prime} - g_X)$ plane in such a scenario is severely restricted by other experiments. 
On top of that, the proposed signal at Belle-II can explore additional regions of the parameter space.

Another important feature emerges from the number of events histogram against the photon energy. We show that, when the correction to muon anomalous magnetic moment 
is given entirely by the $Z^{\prime}$ contribution, an excess in only the highest energy bin is possible. This means that, if an excess is observed in any but the highest 
energy bin, it would be a signature of SUSY particles contributing to the radiative kinetic mixing. This also indicates that an additional source of muon anomalous magnetic 
moment is required beyond the $Z^\prime$ contribution.   

\section{${U(1)_{L_\mu - L_\tau}}$ model in SUSY} 
We extend the MSSM with a new $U(1)$ gauge symmetry, $U(1)_{L_\mu - L_\tau}$, where the (s)muon 
and (s)tau fields with their corresponding (s)neutrinos couple to the additional $Z^{\prime}$ with equal and opposite charge. Because our focus in this work is  mainly on the gauge kinetic mixing, we shall not 
provide much details of the model here. Rather, we concentrate on how the kinetic mixing involving the photon and the $Z^\prime$ appears radiatively in our setup and its 
consequences at the Belle-II experiment.  

\begin{figure}[t]
\includegraphics[width=7cm]{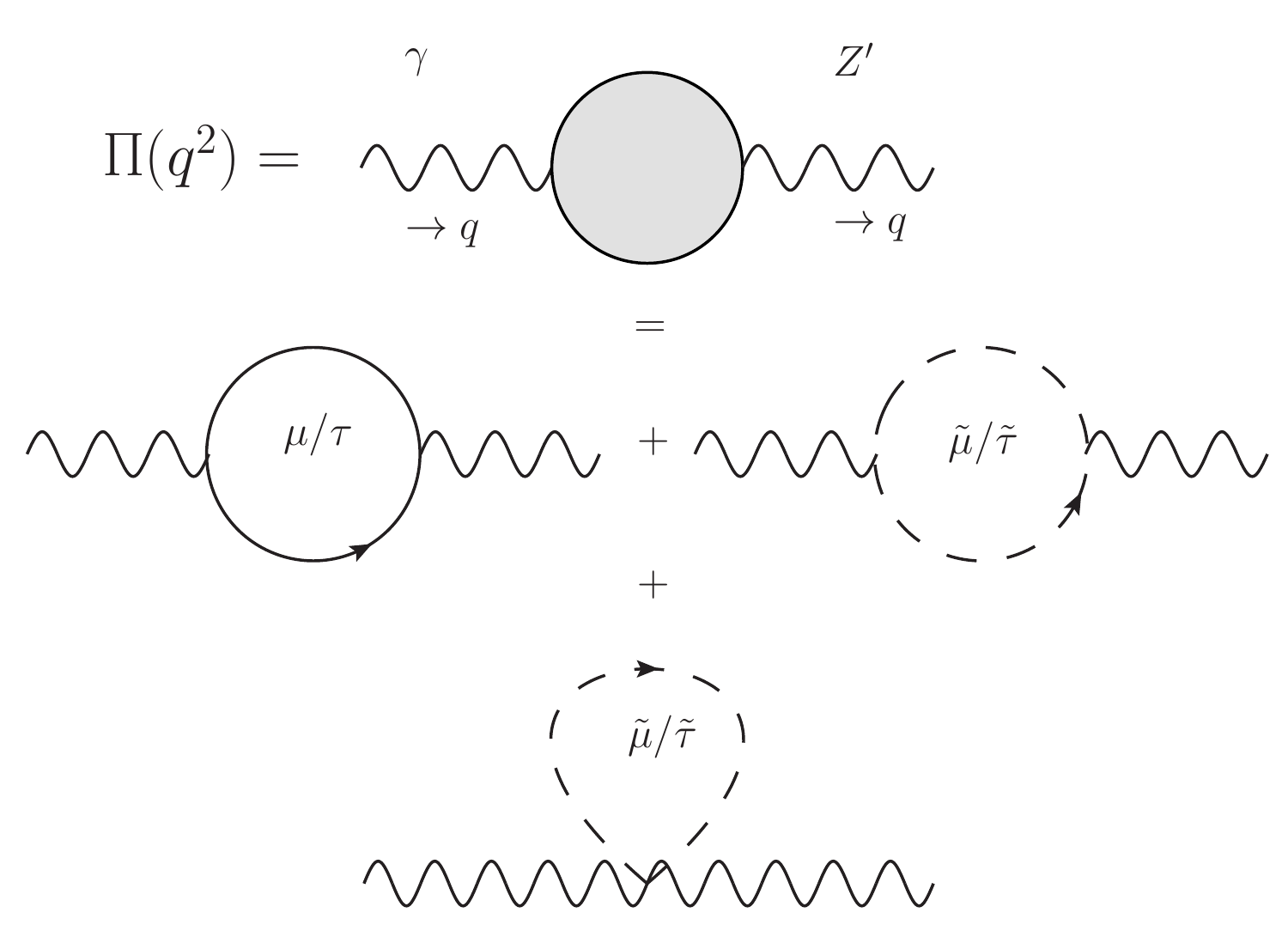}
\caption{Diagrams showing how $\gamma-Z^{\prime}$ kinetic mixing arises radiatively at the one-loop level.}\label{fig:kinetic-mixing}
\end{figure}

We shall assume that the kinetic mixing between $U(1)_{\rm em}$ and \newu given by
\bea
{\cal L}_{kin-mix} &=& \frac{\epsilon}{2}(({\hat W}^{(em) \alpha} {\hat W}^{L_\mu-L_\tau}_\alpha)_F
\eea
is absent at the tree level, i.e., $\epsilon$ = 0. Here ${\hat W}^{(em) \alpha}$ and ${\hat W}^{L_\mu-L_\tau}_\alpha$ are the corresponding SUSY field strengths. 
The subscript ``F" indicates F-term contribution.
However, such a kinetic mixing can still be generated at the one-loop level involving muon, tau, smuon, and stau in the loops as shown in Fig.\ref{fig:kinetic-mixing}. 
The absence of kinetic mixing at the tree level can be justified using some symmetry arguments. One possibility is that the kinetic mixing is forbidden by a discrete 
symmetry, $\mu \leftrightarrow \tau$, ${\tilde \mu} \leftrightarrow {\tilde \tau}$, ${\hat W}^{(em) \alpha} \rightarrow {\hat W}^{(em) \alpha}$, and 
${\hat W}^{L_\mu-L_\tau}_\alpha \rightarrow -{\hat W}^{L_\mu-L_\tau}_\alpha$ in the limit of $m_\mu = m_\tau$ and $m_{\tilde \mu} = m_{\tilde \tau}$. Breaking the 
symmetry softly by $m_\mu \neq m_\tau$ and $m_{\tilde \mu} \neq m_{\tilde \tau}$ generates a finite kinetic mixing radiatively.

Another possibility is to consider the \newu gauge factor embedded within an unbroken nonabelian gauge symmetry (such as $SU(2)$). This would forbid the kinetic mixing. 
However, when the $U(1)$ gauge factor comes out from the breaking of the full nonabelian gauge symmetry, the mass degeneracy of the states within the nonabelian
multiplets will be lost, leading to nonzero kinetic mixing generated at the one-loop level \cite{Dienes:1996zr}. 

As the decoupling theorem does not apply if the heavy (s)particle masses break symmetries \cite{Appelquist:1974tg,Cheng:1997sq}, in this case, nondecoupling 
effects will be present. This will be elaborated in the subsequent discussion. Note also that the superpartner soft masses break supersymmetry and hence may give rise 
to nondecoupling corrections \cite{Cheng:1997sq}.

As discussed, coupling of the photon with  $Z^\prime$ appears at the one-loop level and is given by\footnote{Note that it is sufficient to consider kinetic mixing 
between $U(1)_{\rm em}$ and the $U(1)_{L_\mu - L_\tau}$ as long as $M^2_{Z^\prime}/M^2_Z \ll 1$.} 
\bea
\epsilon \equiv \Pi(q^2) &=& \frac{8 e g_X}{(4 \pi)^2} \int_0^1 x(1-x)\ln\frac{m^2_\tau - x(1-x)q^2}{m^2_\mu - x(1-x)q^2}dx \nn
+&&\hspace{-1.7em} \frac{2 e g_X}{(4 \pi)^2} \int_0^1 (1-2x)^2 \ln\frac{m^2_{\tilde \tau} - x(1-x)q^2}{m^2_{\tilde \mu} - x(1-x)q^2}dx
\eea 
where the contributions come from the loop diagrams in Fig.\ref{fig:kinetic-mixing}.
Here $e$ is the electromagnetic charge, $m_\ell$ and $m_{\tilde \ell}$ are the masses of charged lepton $\ell$ and charged slepton ${\tilde \ell}$,
$q$ is the momentum carried by $\gamma$ and $Z^\prime$ and $g_X$ is the gauge coupling corresponding to $U(1)_{L_\mu - L_\tau}$. Here, for simplicity of the 
analysis, we have considered identical masses for $m_{{\tilde \ell}_L}$ and $m_{{\tilde \ell}_R}$. In general these masses can be different and in such cases 
one must define other ratios involving only left-handed sleptons or right-handed sleptons. 

\section{Constraints and signature at Belle-II}
At the Belle-II experiment we shall consider the signal process $e^+e^- \rightarrow \gamma Z^\prime$ and then $Z^\prime$ 
decaying to $\nu {\bar \nu}$ leading to a final state $e^+e^- \rightarrow \gamma + \mE.$ The kinetic mixing parameter is a function of $q^2 = M^2_{Z^\prime}$ (for on-shell production of the $Z^{\prime}$ boson) and 
$r$. It is the dependence on $r$ that makes the model predictions very different compared to what is obtained 
in gauged $L_\mu - L_\tau$ models with no supersymmetry. In our analysis the kinetic mixing parameter, $\epsilon$, that is generated radiatively, never exceeds $10^{-4}$. 

At this stage it is worth mentioning that the gauged $L_\mu -L_\tau$ model was first introduced to address the discrepancy between the experimental measurement
and the SM predictions of muon $(g-2)$, and this is given by\cite{Olive:2016xmw} 
\bea
a_\mu^{\rm exp} - a_\mu^{\rm SM} = (28.7 \pm 8.0) \times 10^{-10} 
\eea
where $a_\mu \equiv (g_\mu-2)/2.$ 

In gauged $L_\mu - L_\tau$ models without supersymmetry, the extra contribution to $(g_\mu-2)/2$ comes solely from a one-loop diagram involving $Z^\prime$ 
and is given by\cite{Baek:2001kca,Ma:2001md}
\bea
a_\mu^{Z^\prime} = \frac{g^2_X}{8 \pi^2}\int_0^1 \frac{2 m^2_\mu x^2 (1-x)}{x^2 m^2_\mu + (1-x)M^2_{Z^\prime}} dx.
\eea
In addition, the most stringent constraints on the \newu parameter space come from neutrino trident production 
(CCFR\cite{Mishra:1991bv}), neutrino-electron elastic scattering (BOREXINO\cite{Bellini:2013lnn}) and the light $Z^\prime$ search through 
$e^+e^- \rightarrow \mu^+ \mu^- Z^\prime, ~ Z^\prime \rightarrow \mu^+ \mu^-$ by the BaBar 
Collaboration \cite{TheBABAR:2016rlg}. The CCFR collaboration reported a strong adherence of the observed cross section to the SM prediction,
which strongly constrains a large section of the $M_{Z^{\prime}}$-$g_X$ parameter space \cite{Altmannshofer:2014pba}. 
The observation of $^7$Be solar neutrino scattering rates at Borexino disfavors any additional contribution that is $8\%$ or more above the 
SM prediction\cite{Harnik:2012ni,Kamada:2015era}. For a recent discussion on other constraints, see, Ref.\cite{Bauer:2018onh}.
Taking them into account, a thin slice of the $M_{Z^{\prime}}$-$g_X$ parameter space,
\bea
\text{10 MeV}\lesssim M_{Z^{\prime}}\lesssim\text{210 MeV}, ~4\times 10^{-4}\lesssim g_X\lesssim 10^{-3}\label{eqn:rangens}
\eea
is left to explain the muon $(g-2)$ anomaly.
However, it was shown in \cite{Banerjee:2018eaf} that once SUSY is taken into consideration
the allowed parameter range satisfying muon ($g-2$) is larger depending on the choice
of SUSY parameters. 

In case of superheavy sparticles, the allowed parameter range is the same as that in non-SUSY $L_\mu - L_\tau$ [given in Eq.(\ref{eqn:rangens})]; however, their contribution
to kinetic mixing is nondecoupling. Hence, one would still be able to discern their signatures at Belle-II through the signal process.

\begin{figure}[t]
\centering
\subfigure[\label{fig:CS-diagram}]{\includegraphics[width=6cm]{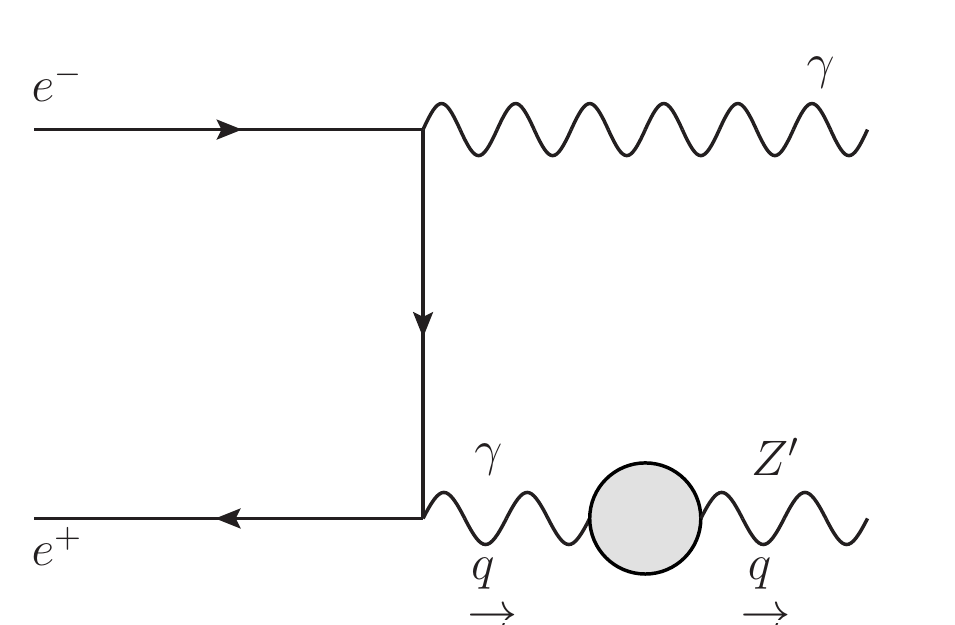}}
\subfigure[\label{fig:cross-section}]{\includegraphics[width=8cm]{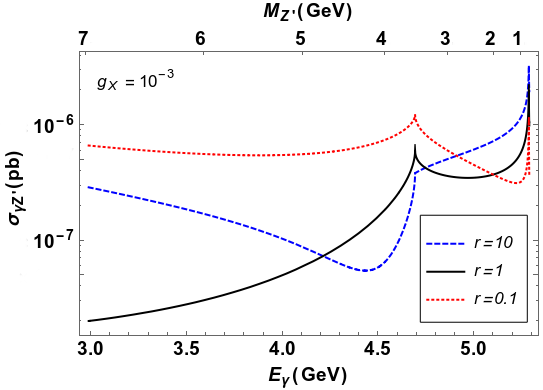}}
\caption{(a) Feynman diagram for $\gamma-Z^{\prime}$ production at Belle-II. (b) Variation of the cross section for this process with changing $E_{\gamma}$ and $M_{Z^{\prime}}$.}\label{fig:sigproc}
\end{figure}
\begin{figure*}
\centering
\subfigure[\label{fig:hg-3-10}]{\includegraphics[width=5.9cm]{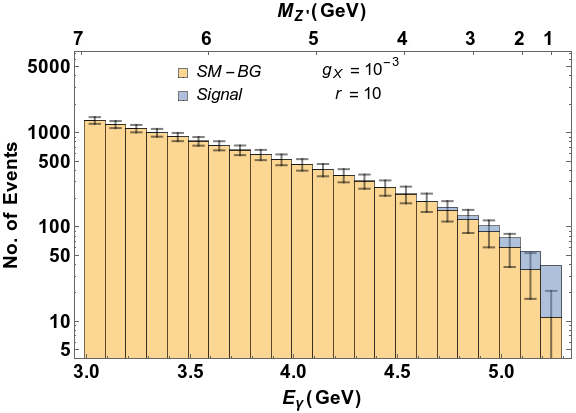}}
\subfigure[\label{fig:hg-3-1}]{\includegraphics[width=5.9cm]{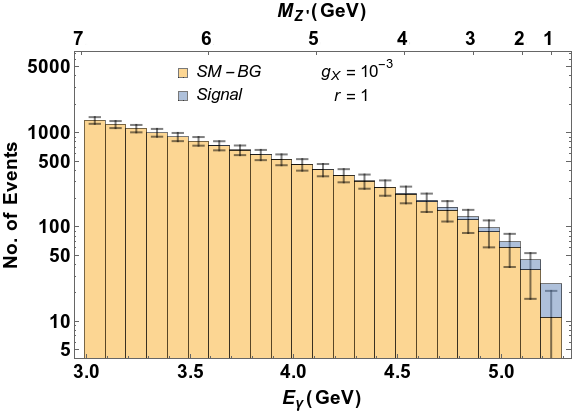}}
\subfigure[\label{fig:hg-3-0-1}]{\includegraphics[width=5.9cm]{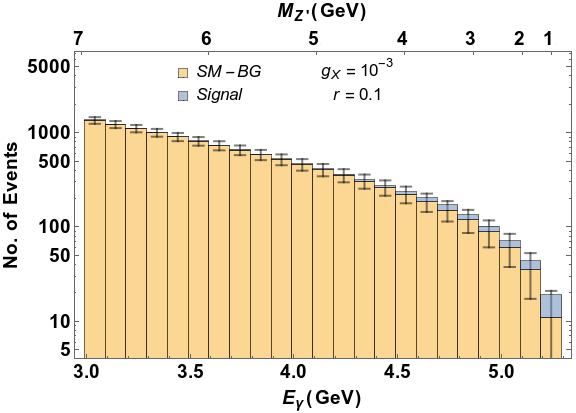}}
\subfigure[\label{fig:hg4-3-10}]{\includegraphics[width=5.9cm]{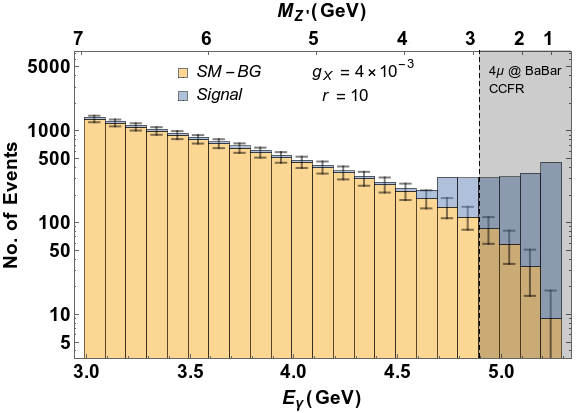}}
\subfigure[\label{fig:hg4-3-1}]{\includegraphics[width=5.9cm]{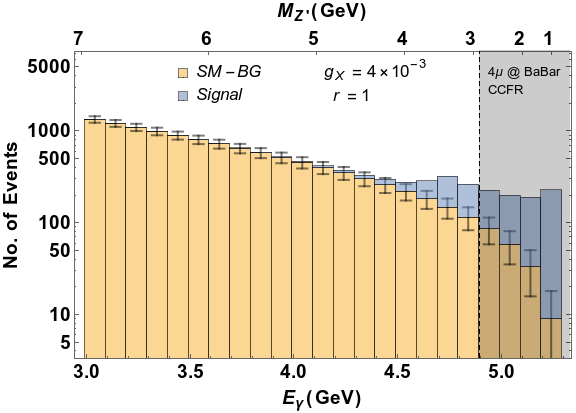}}
\subfigure[\label{fig:hg4-3-0-1}]{\includegraphics[width=5.9cm]{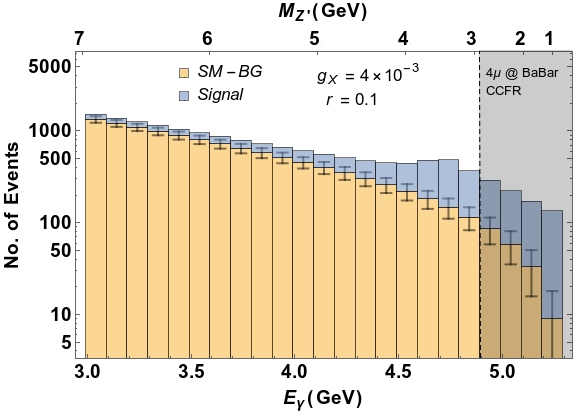}}
\caption{$E_\gamma$ distribution of event numbers in SUSY $L_\mu - L_\tau$ model for $e^+e^-\rightarrow \gamma \mE$ channel at the Belle-II experiment with $\sqrt{s}=10.58$ GeV and an integrated luminosity of 50 ab$^{-1}$. The shaded regions are excluded from the observations of the CCFR \cite{Mishra:1991bv} and BaBar \cite{TheBABAR:2016rlg} collaborations.}\label{fig:hist}
\end{figure*}
Belle-II experiment \cite{Abe:2010gxa,Kou:2018nap} is an electron-positron collider with a center-of-mass energy of $\sqrt{s}$ = 10.58 GeV and is expected to reach an 
integrated luminosity of 50 ${\rm ab}^{-1}.$ The signal process $e^+e^- \rightarrow \gamma + \mE$ under study results from the associated production of a monoenergetic photon and a 
light $Z^\prime$ [see, Fig.\ref{fig:CS-diagram}] and subsequent decay of $Z^\prime$ into a $\nu {\bar \nu}$ pair. The $\gamma-Z^\prime$ kinetic mixing depends on the 
momentum $q$ carried by $Z^\prime$ as well as the ratio of smuon and stau masses. The cross section of $Z^\prime$ production $e^+e^- \rightarrow \gamma +  Z^\prime$ in 
the center-of-mass frame is given by \cite{Essig:2009nc}
\bea
&&\sigma(e^+e^- \rightarrow \gamma +  Z^\prime) = \frac{2 \pi \alpha^2 |\Pi(M^2_{Z^\prime})|^2}{s} \bigg[ 1- \frac{M^2_{Z^\prime}}{s} \bigg] \nonumber \\
&\times& \bigg[\bigg \{1 + \frac{2sM^2_{Z^\prime}}{(s-M^2_{z^\prime})^2}\bigg \}\ln\frac{(1 + \cos\theta_{\rm max})(1 - \cos\theta_{\rm min})}{(1 - \cos\theta_{\rm max})
(1 + \cos\theta_{\rm min})} \nonumber \\
&-&\cos\theta_{\rm max} + \cos\theta_{\rm min} \bigg].
\eea
Here we have $\cos\theta_{\rm min} <\cos\theta < \cos\theta_{\rm max}$, with $\cos\theta_{\rm min} = -0.821$ and $\cos\theta_{\rm max} = 0.941$, which corresponds to the range of the
coverage of the electromagnetic calorimeter \cite{Araki:2017wyg}. The angle $\theta$ is the angle between the electron beam axis and the photon momentum. 
The cross section is plotted in Fig.\ref{fig:cross-section}, 
where the final state photon energy $E_\gamma$ is related to $q^2$ in the center-of-mass frame as
\bea
E_\gamma = \frac{s - q^2}{2 \sqrt{s}} .\label{eqn:eg-mzp}
\eea 
The maximum value of $E_{\gamma}$ is $\sqrt{s}/2$ (5.29 GeV at Belle-II) that corresponds to $M_{Z^{\prime}}^2=0$ for an on-shell $Z^{\prime}$. At Belle-II this process can probe the $Z^\prime$ gauge boson of mass $\lesssim 6$ GeV, which corresponds to a maximum $g_X$ of $4\times 10^{-3}$ \cite{Sirunyan:2018nnz}. The decay mode of the $Z^{\prime}$ boson into two muons is possible for $M_{Z^\prime}> 2m_{\mu}$ and results in $\gamma\mu^+\mu^-$ signal which is cleaner. However it cannot probe the crucially important range of $Z^{\prime}$ mass that can still explain muon $(g-2)$ in the absence of additional SUSY contribution. 

The decay width of the additional gauge boson is much 
less than its mass for this parameter space, which justifies the use of the narrow width approximation.
The value of the gauge coupling $g_X$ has been taken to be $10^{-3}$ in Fig.\ref{fig:cross-section} to correspond to a region where muon $(g-2)$ may still be satisfied even when the superpartners 
are very heavy. 

One can see from this figure that the cross section increases for higher values of $E_\gamma$, which corresponds to lower $M_{Z^\prime}$. 
The $r=1$ curve corresponds to the case where the slepton masses are equal and hence their contribution to the radiative kinetic mixing
drops out. We consider a hierarchy between the sleptons to the tune of a factor of 10 at its maximum extent. It is not 
difficult to translate this choice into a realistic situation where the lighter of the two sleptons are at about 1 TeV whereas the other is close to 10 TeV. The two 
different ratios, 0.1 and 10, then represent two very different phenomenological situations, one where the stau is much lighter than the smuon and the other the opposite, 
respectively.
In Fig.\ref{fig:cross-section}, a heavier stau mass ($r>1$) leads to a larger cross section at higher photon energies. 
The charged slepton contribution to $\epsilon$ interferes destructively with the charged lepton contribution when $r<1$ in this region, resulting in a smaller 
cross section. However the same may not be said for the lower photon energies. In fact, for a large part of the photon energy spectrum, a lighter stau mass results in 
a larger cross section.

The SM background comes from the 2 $\rightarrow$ 3 process $e^+ e^- \rightarrow \gamma \nu {\bar \nu}$ involving $W$ and $Z$ bosons in the propagator. The differential scattering cross section in the center-of-mass frame can be found in Ref.\cite{Araki:2017wyg}. We have assumed 
that the electromagnetic calorimeter does not miss any event if the photon is within the detector range. We also do not consider the situation where one of the photons 
in a $\gamma\gamma$ final state or an $e^+e^-$ pair in radiative Bhabha scattering is missed and contributes to the background.

We have compared the number of events corresponding to the signal and the background processes in Fig. \ref{fig:hist} for an integrated luminosity 
of 50~${\rm ab}^{-1}$. In these plots the width of each energy bin is taken to be $\Delta E_\gamma = 0.1 ~{\rm GeV}$ \cite{Abe:2010gxa},
which is the detector resolution for the photon energy at Belle-II. The values of the coupling 
$g_X$ have been taken to be $10^{-3}$ and $4 \times 10^{-3}$. The choice $4 \times 10^{-3}$ is consistent with the latest measurement of $Z \rightarrow 4 \mu$ at the 
LHC \cite{Sirunyan:2018nnz} as well as the results from the light $Z^\prime$ search by the BaBar Collaboration for $M_{Z^\prime} \gtrsim$ 3 GeV \cite{TheBABAR:2016rlg}. 
However, it requires additional SUSY contribution to the anomalous magnetic moment of the muon. The ratio $r$ is allowed to take the values 0.1, 1 and 10. The statistical 
errors are estimated using $\sqrt{N_S + N_B}$ where $N_S$ and $N_B$ denote the number of signal events and background events respectively. As we can see from the 
Fig.\ref{fig:hg-3-10}-\ref{fig:hg-3-0-1}, in the case of $r = 1$, i.e. when SUSY gives no contribution, the significance ($\equiv N_S/\sqrt{N_S + N_B}$) of the signal 
events compared to the SM background is larger than 3$\sigma$ only in the highest energy bin for $g_X=10^{-3}$. On the other hand, for a larger ratio, i.e. $r = 10$, 
the 3$\sigma$ excess can be seen in the last two bins with an excess of around 5$\sigma$ in the last bin, which is clearly different from the $r = 1$ case. However, no 
significant excess is observed in any of the bins for a smaller ratio, $r=0.1$, at this value of $g_X$.
\begin{figure}[b]
\centering
\vspace*{-1.0in}
\includegraphics[width=7cm]{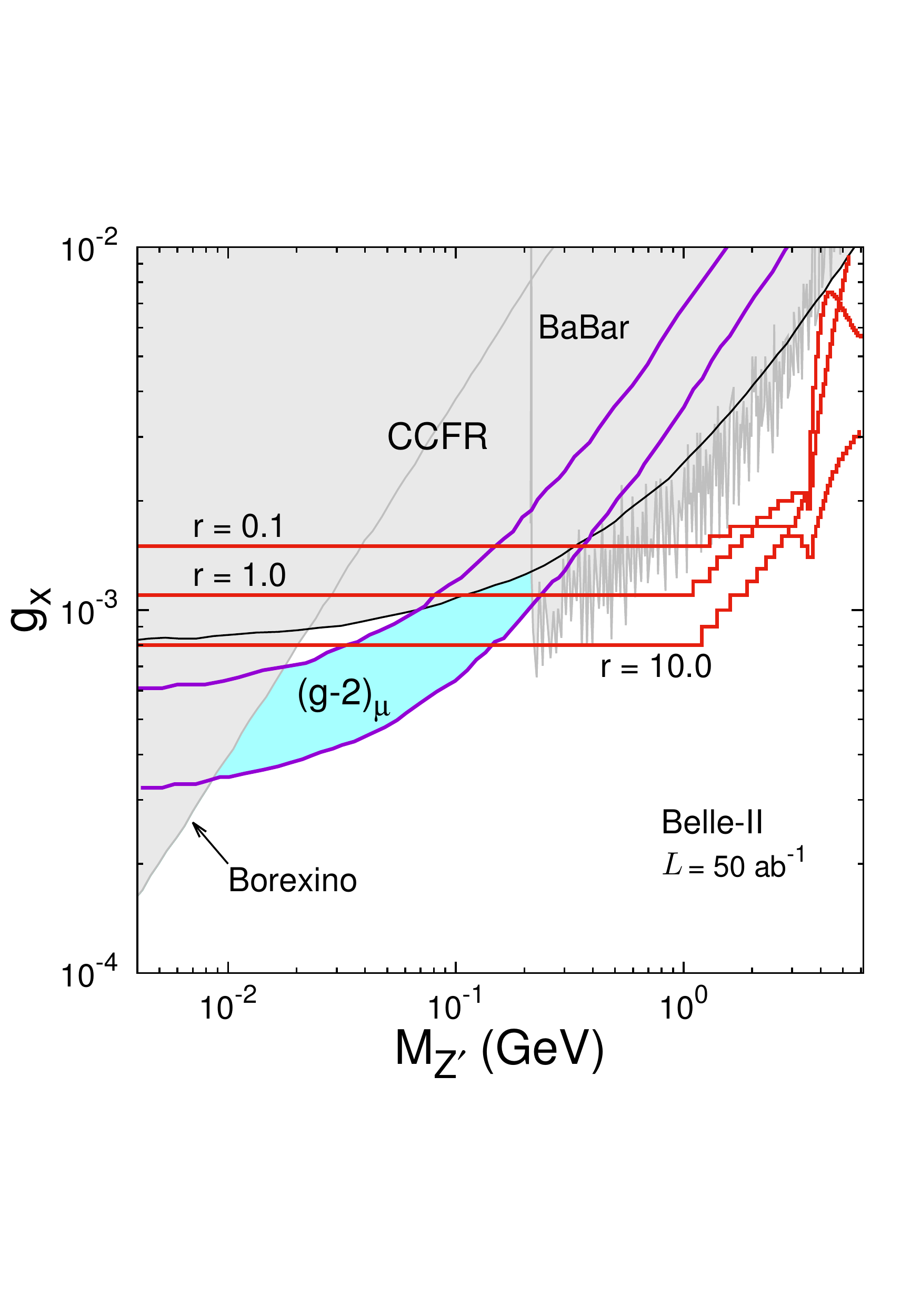}
\vspace*{-0.65in}
\caption{3$\sigma$ exclusion plot in $M_{Z^\prime} - g_X$ plane. Larger values of $r$ exclude a larger region and vice versa. The exclusion region from BOREXINO is shown for $r$ = 1 
and depends very mildly on this ratio: a larger $r$ excludes a larger region. Allowed region at 2$\sigma$ from muon magnetic moment measurements is shown when the sparticles do not contribute to $(g-2)_{\mu}$.}\label{fig:contour-mzprime-gx}
\end{figure}
The result is more spectacular in the case of $g_X = 4 \times 10^{-3}$ as seen in Figs.\ref{fig:hg4-3-10}-\ref{fig:hg4-3-0-1}. We get more than 3$\sigma$ excess in a large 
number of bins for all three choices of $r$, although the last four energy bins are ruled out in this case from the observations of the CCFR and BaBar Collaborations. These histograms are even more intriguing when analyzed from the vantage point of muon $(g-2)$.
Note that any value of $M_{Z^\prime}$ less than 1.38 GeV corresponds to $E_{\gamma} > 5.2$ GeV from Eq.(\ref{eqn:eg-mzp}). Hence, Eq.(\ref{eqn:rangens}) restricts any observable excess to be only in the highest energy bin, irrespective of $r$, when all the sparticles are too heavy to contribute to $(g-2)_{\mu}$. 
This additional information, in conjunction with the result of Ref.\cite{Banerjee:2018eaf}, which shows that higher values of both $M_{Z^{\prime}}$ and $g_X$ may be allowed 
in a supersymmetric model, results in a very important inference. If Belle-II observes any significant excess in any of the energy bins apart from the last one, it would be 
an unmistakable signature of the sleptons contributing to the $\gamma-Z^{\prime}$ kinetic mixing\footnote{Even when r = 1 i.e. when
the sleptons do not contribute to the kinetic mixing, such an excess is
impossible in a non-SUSY scenario if the muon (g − 2) observations are
to be explained.} and an additional source of $(g-2)_{\mu}$ over the $Z^{\prime}$ 
contribution. 

An excess in the last energy bin, while still a signature of \newu, can be present even in the absence of SUSY. In that case the maximum possible excess 
in the last bin comes for $g_X = 10^{-3}$. We can then define a ratio $N_S/N_0$ where $N_S$ is the number of signal events observed and $N_0$ is the maximum 
number of events in the absence of SUSY (i.e., for $g_X = 10^{-3}$). If a significant excess is observed in the last bin and this ratio is found to be 
greater than 1 then it would be a telltale signature of the presence of SUSY. In a supersymmetric scenario, this ratio depends on both $r$ and $g_X$ and an 
observation would point towards a range of allowed values for them. An ambiguity would persist if this observed ratio is less than 1, although that would 
constrain the non-SUSY $L_\mu -L_\tau$ model very strongly ($9\times 10^{-4}\lesssim g_X \lesssim 10^{-3}$). 

\begin{figure}[t]
\centering
\includegraphics[width=7cm]{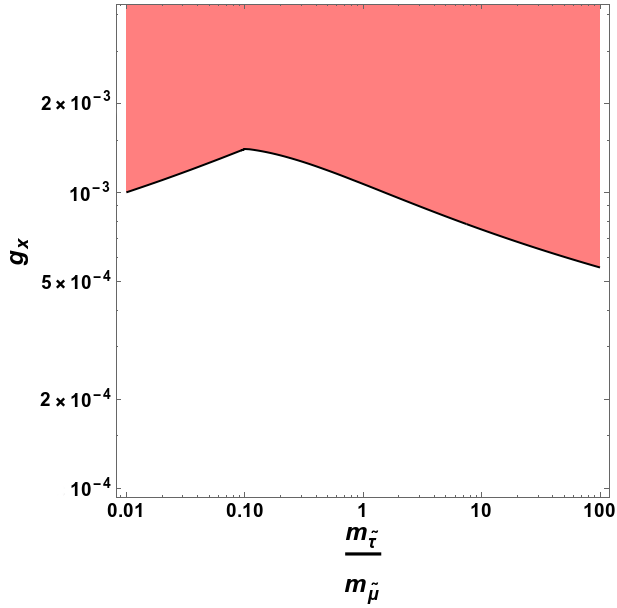}
\caption{3$\sigma$ exclusion plot in the $g_X - r$ plane for $M_{Z^{\prime}}<7$ GeV. Values of $g_X$ above $4\times 10^{-3}$ are disallowed for this range in gauge boson mass from the 
constraints discussed in the text.}\label{fig:contour-gx-r}
\end{figure}
The absence of any observed excess in this channel would exclude regions of the $M_{Z^{\prime}}-g_X$ parameter space that are still allowed to explain muon $(g-2)$, 
as shown in Fig.\ref{fig:contour-mzprime-gx}. This exclusion region would depend on the ratio $r$ and supersedes any existing bounds for much of the parameter space. 
We show the exclusion contours for $r$ = 0.1 and 10 which makes the variation of the excluded region with $r$ very clear. Parts of the additional unconstrained region 
in this plane, which may be available to explain $(g-2)_{\mu}$ depending on SUSY parameters, are also excluded by the lack of observed excesses in the signal channel. 
As shown in Fig.\ref{fig:contour-gx-r}, lack of any significant excess would also strongly constrain the hitherto free $g_X-r$ parameter space. This exclusion region 
corresponds only to values of $Z^{\prime}$ mass less than 7 GeV. Heavier $Z^{\prime}$ gauge boson parameter space would still remain unconstrained, however, explaining 
muon $(g-2)$ in such a scenario would require significant SUSY contribution.  

\section{Conclusion}
We have considered the possibility of an additional $L_{\mu}-L_{\tau}$ force within a supersymmetric framework. We demonstrated the exciting nondecoupling behavior of the contribution of sleptons to the $\gamma-Z^{\prime}$ kinetic mixing in this class of models. This important observation begets the possibility of visible signatures of scalars like the smuon and stau, too heavy to have been detected at the LHC, in processes modified by the $\gamma - Z^{\prime}$ mixing.  We propose the channel $e^+e^-\rightarrow\gamma Z^{\prime}\rightarrow\gamma\mE$ to probe this effect and show that the Belle-II is ideally equipped to do so. 

It is clear that, if Belle-II observes an excess beyond $3\sigma$ in any of 
the energy bins except the highest one, it would be an undeniable signature of heavy charged scalars, like those appearing in SUSY, contributing to the 
kinetic mixing. It would also indicate the need for additional contribution, beyond that from $Z^{\prime}$, to accommodate muon anomalous magnetic moment measurements. Finally, if Belle-II observes no excess whatsoever over the SM in any of the energy bins, large chunks of both 
$M_{Z^{\prime}}-g_X$ and $g_X-r$ parameter space would be excluded. Additionally, this feature and its analysis would remain unchanged for any model with an extra 
gauged $U(1)$ symmetry incorporating chiral superfields that have equal and opposite charge under it.

\begin{acknowledgments}
We thank Satyanarayan Mukhopadhyay for many helpful discussions and a careful reading of the manuscript. We also thank Anirban Kundu and 
Subhadeep Mondal for helpful discussions.
\end{acknowledgments}

\end{document}